\documentclass[sigconf]{acmart}
\usepackage{graphicx,wrapfig,lipsum}
\AtBeginDocument{%
  \providecommand\BibTeX{{%
    \normalfont B\kern-0.5em{\scshape i\kern-0.25em b}\kern-0.8em\TeX}}}

\acmConference[arxiv.org]{arxiv.org}{October}{2020}

\setcopyright{none}
\renewcommand\footnotetextcopyrightpermission[1]{}
\begin{document}

\title[Fat Pad Cages for Facial Posing]{Fat Pad Cages for Facial Posing}

\author{Adèle Colas}
\affiliation{%
  \institution{IMT Atlantique, France}
}
\email{adele.colas@telecom-bretagne.eu}

\author{Florent Guiotte}
\affiliation{%
  \institution{ESIR, France}
}
\email{florent.guiotte@uhb.fr}

\author{Fabien Danieau}
\affiliation{%
  \institution{InterDigital, France}  
}
\email{fabien.danieau@interdigital.com}

\author{François Le Clerc}
\affiliation{%
  \institution{InterDigital, France}  
}
\email{francois.leclerc@interdigital.com}

\author{Quentin Avril}
\affiliation{%
  \institution{InterDigital, France}  
}
\email{quentin.avril@interdigital.com}

\renewcommand{\shortauthors}{Colas et al.}

\begin{abstract}
We introduce Fat Pad cages for posing facial meshes. It combines cage representation and facial anatomical elements, and enables users with no artistic skill to quickly sketch realistic facial expressions. The model relies on one or several cage(s) that deform(s) the mesh following the human fat pads map. We propose a new function to filter Green Coordinates using geodesic distances preventing global deformation while ensuring smooth deformations at the borders. Lips, nostrils and eyelids are processed slightly differently to allow folding up and opening. Cages are automatically created and fit any new unknown facial mesh. To validate our approach, we present a user study comparing our Fat Pad cages to regular Green Coordinates. Results show that Fat Pad cages bring a significant improvement in reproducing existing facial expressions. 
\end{abstract}

%
%

\begin{teaserfigure}
  \includegraphics[width=\textwidth]{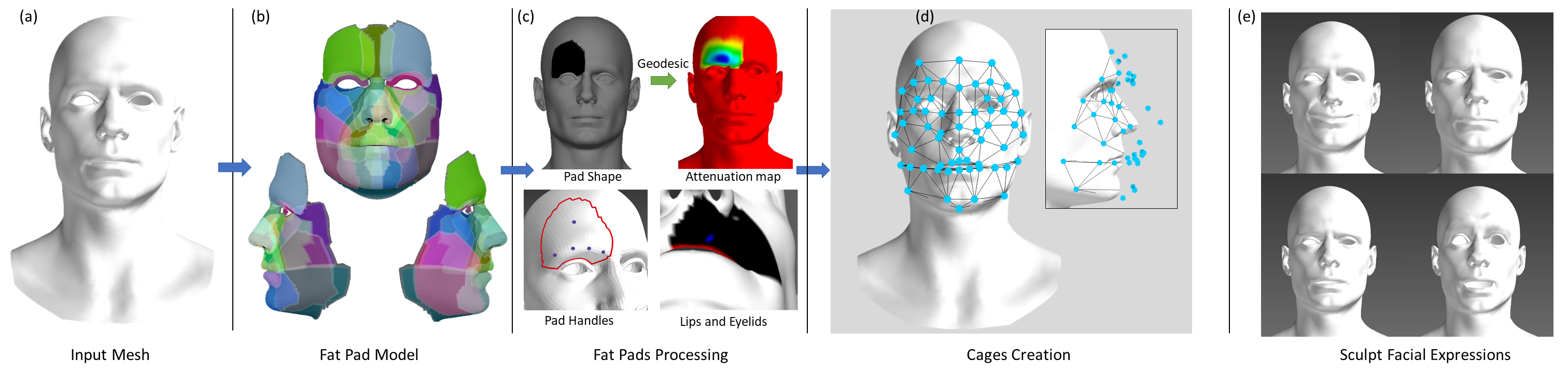}
  \caption{Given any facial mesh~(a), fat pads are computed following a template fat map~(b), attenuation matrices are computed using geodesic distances~(c top), pad's handles are mapped onto the mesh~(c bottom) and a cage is built using convex hull of Delaunay triangulation~(d). This automatically generated Fat Pad cage enables to easily and quickly sculpt facial poses~(e).}
  \label{fig:teaser}
\end{teaserfigure}

\maketitle

\section{Introduction}

Facial expressions are key elements in the realistic characters creation process \cite{orvalho2012facial}. Recent movie characters have facial rigs with thousands of blend shapes. Blend shapes represent facial elementary displacements and are often related to muscles \cite{Joshi2003}. Modeling them requires strong modeling skills and a lot of time (weeks for hero characters). Posing faces can be achieved through linear piece-wise modeling approaches \cite{Tena2011} that require a lot of facial performances motion capture data. Physics-based approaches are also possible~\cite{Ichim2017} but a high-end polyhedral facial anatomical representation is necessary to simulate bones, muscles and fat. Simpler solution would use any free-form deformation (FFD) techniques \cite{Sorkine2004}. However, FFD are generic and do not take into account facial semantics. They can lead to severe uncanny artifacts for non-experts as a face is deformed as any other meshes. These non-expert users may be professionals, like stage directors who want to sketch facial expressions on a tablet to give guidelines, or beginners such as digital art school students.
The final shape of a facial expression is a complex combination of the skull, the muscles and the fat~\cite{uldis2017}. As skin is soft and thin, it may rather be considered as a protective and aesthetic layer than a dynamic and active one. The skull gives the macroscopic shape of the face, and except for the jaws, it should not move. Actually, muscles are the cause of the action, and fat is the main visual element of the final shape (wrinkles, dimples, gaps and folds). Fat on the face is divided in several pads. The name "fat pad" comes from medical dictionaries and plastic surgery. Each pad moves almost independently if it is activated by the related muscle, and the skin joins all the pads together. We present the novel concept of fat pads for facial posing, enabling an interactive mesh deformation that respects the constraints of a face and is intuitive and easy to use for beginners. We propose a new automatic way for creating facial cages and a new weighted function based on geodesic distances to localize and limit the influence of cage deformation. Finally, we present a user study with 17 naive users that validates the easiness and intuitiveness of the Fat Pad cages.

\section{Related work}

\textbf{Cages.} Cages are polyhedra with a small number of polygons encompassing a mesh. Mesh vertices are linear sums of the cage's vertices multiplied by weight functions (coordinates). Floater et al.\ \cite{Floater2003} first proposed a way to use cages with a random geometry with the Mean Values Coordinates (MVC). But undesired artifacts appear with negative coordinates. Joshi et al.\ then proposed the Harmonic Coordinates (HC) that resolve Floater's issues but require a long computation time~\cite{Joshi2007}. Lipman et al.\ put forward the Green Coordinates (GC) that do not have negativity issues~\cite{Lipman2008}. Computation time allows interactivity, surface's details and mesh volume are correctly preserved. Jacobson et al.\ propose three rules for cage generation~\cite{Jacobson2014}: it must be low resolution to reduce manual interactions, it must fully and tightly bind the enveloped model and it should respect the mesh topology for the manipulation to be intuitive. Yang et al.\ propose to use a template based cage generation type~\cite{Yang2013}. The nested cage~\cite{Sacht2015} uses the character mesh to build progressive decimated cage mesh around the model. Xian proposes to generate oriented bounding boxes for each mesh part and then registering them together~\cite{Xian2012} but it is hard to apply on faces due to the lack of articulation. Chen et al.\ proposes a skeleton based approach~\cite{Chen2014} but face meshes do not have any skeleton. Lee et al.'s cage relies on user inputs~\cite{Lee1995} that cuts the model and the slides set is used to construct an initial cage, but our target is an automatic approach with no user interaction.

\textbf{Facial Posing.} To model faces, there are several approaches such as blend shapes~\cite{Joshi2003, lewis2010, seo2011, lewis2014}, region-based model~\cite{Tena2011} and physics-based model~\cite{SifakisNF05, CongBF16, Ichim2017}. Blend shapes remain a predominant technique and are in most digital content creation tools. Tena et al.~\cite{Tena2011} propose a region-based linear model, that uses PCA. It separates the mesh in several models and the formulation restricts the solutions to have semi-consistent boundaries while enforcing user-given constraints. It does not allow to model other expressions than the ones from the training set. It is also possible to rely on eigenbases of the Laplacian to generate low-frequency bases~\cite{BouazizWP13} to track human face using RGB-D camera. But it is not compatible with static facial mesh as our input. Waters proposed a muscle model that uses muscle vectors and radial functions derived from linear and sphincter muscles to deform a skin mesh~\cite{Waters1987}. Lee et al.\ presented an algorithm that automatically constructs functional models of heads, subjected to laser-scanned range and reflectance data~\cite{Lee1995}. Contractile muscles within a dynamic skin model are then inserted and rooted in an estimated skull structure with a hinged jaw. Kähler et al.~\cite{Kahler2001} proposed a model for muscle-based facial animation composed of three layers: skin with fatty tissue, muscles attached to the skull, and the underlying bone structure, composed of immovable skull and rotating jaw. Muscle-based models are accurate but not interactive. Ichim et al.\ have presented a physics-based approach to model faces~\cite{Ichim2017} that optimizes facial physical interactions and prevents undesired artifacts, but it requires a complete facial anatomical model.

This highlights a lack between intuitive cage deformers, not used to pose faces, and dedicated professional tools, requiring time and strong artistic skills. Fat Pad cages enable to fill this gap proposing an intuitive, interactive and accurate modeling approach.

\section{Fat Pads}
We propose to divide a face in smaller areas called "fat pads". On each pad, some vertices are manually labeled as "handles" and used to build the cage. Their displacement induces smooth deformations of the corresponding area. This method represents the way facial muscles would deform the skin. The fat pads approach limits the deformation, induced by the cage, to a local area. We use Green Coordinates (GC)~\cite{Lipman2008} but the concept is applicable to any coordinates system~\cite{Joshi2007, nieto2013cage}. We use GC because they better preserve mesh details, are computed at interactive time and the cage does not need to entirely contain the mesh. However, cage's handles only affect the vertices of their own fat pad. To ensure a smooth and continuous deformation between pads, it must be attenuated at the border. Hence, a weight between 0 and 1 smooths the GC. It is associated to each vertex according to its position in a pad. While the vertices close to the handle move normally, the ones close to the border area remain still or almost still.

\textbf{Template Map.} Fat pads are designed following a reference book used by modelers in VFX studios~\cite{uldis2017}. As Cong et al. proposed to simulate flesh and muscles \cite{CongBEBF15}, we use a template map where we manually painted fat pads. This operation has only to be performed once. The template fat pad map is then used for all new meshes. If a character requires specific fat motion behaviors, the template is easily re-paintable and handles freely movable. The face has smaller fat pads or tissues in-between main fat pads. To avoid a complex modeling and fill the gaps between the pads, we propose to overlap them as shown on Figure~\ref{fig:teaser}. It avoids sharp borders and ensure smooth borders' deformations. A plugin in Maya was developed to paint fat pads on meshes and place handles.

\begin{figure}[h]
\centering
\includegraphics[width=\linewidth]{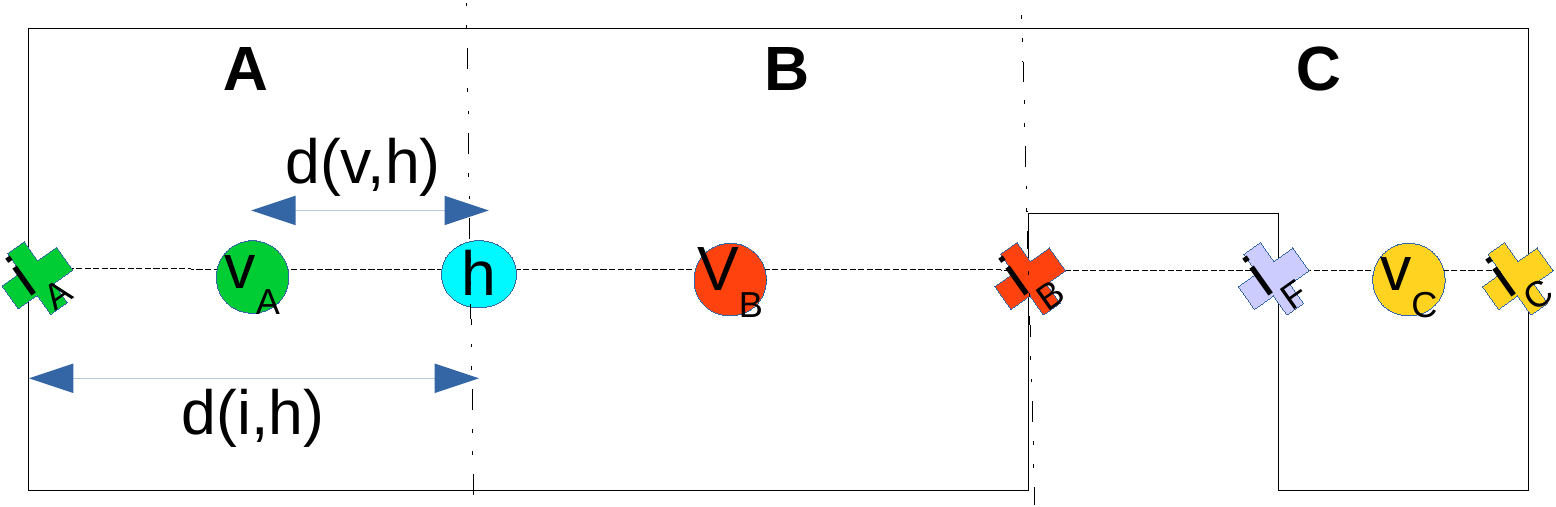}
\caption{A: vertex $v_A$ with handle $h$. $d(i_A,h)$, geodesic dist. between $h$ and $i_A$ intersection between pad's border and $\vec{hv_a}$. B, C: non-convex fat pad, many intersections may exist.}
\label{fig:scheme}
\end{figure}

\begin{figure}[h]
\centering
\includegraphics[height=2.1cm]{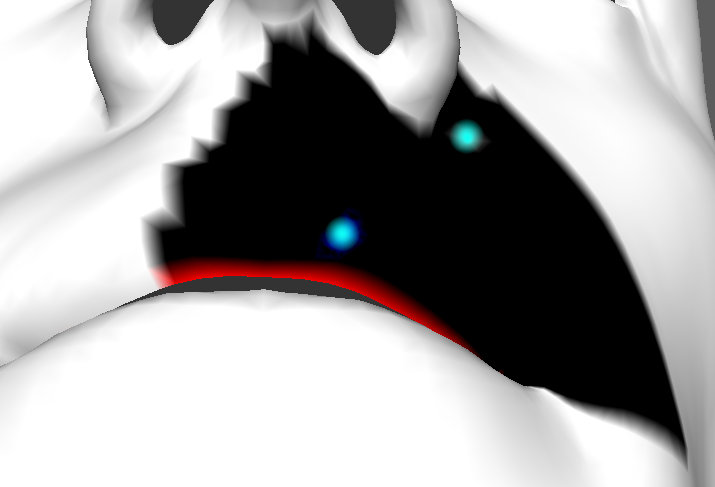}
\includegraphics[height=2.1cm]{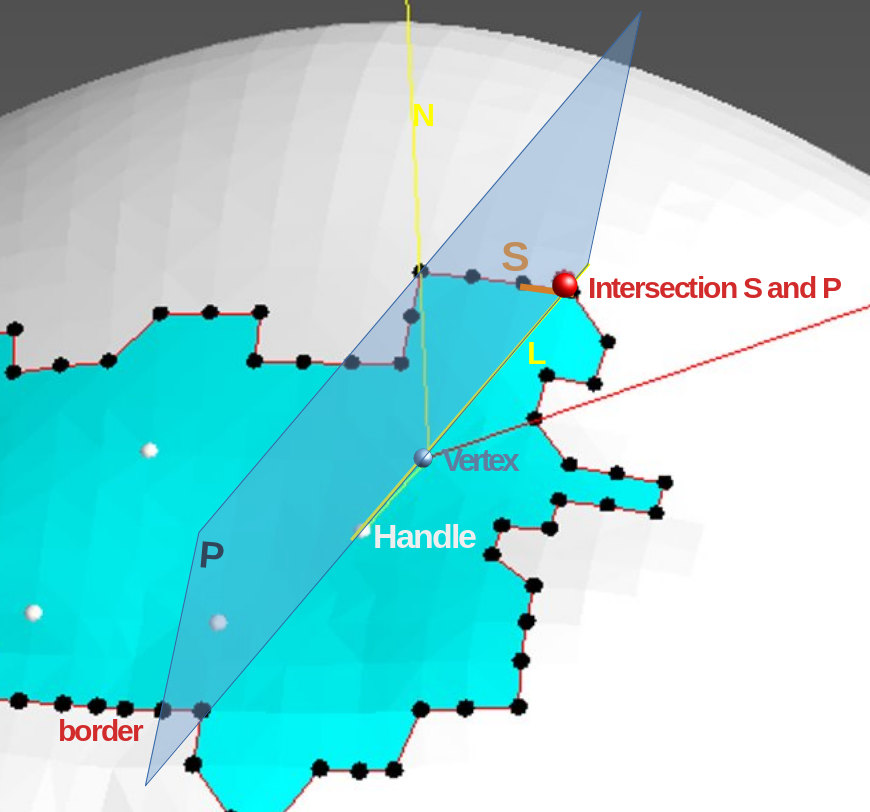}
\includegraphics[height=2.1cm]{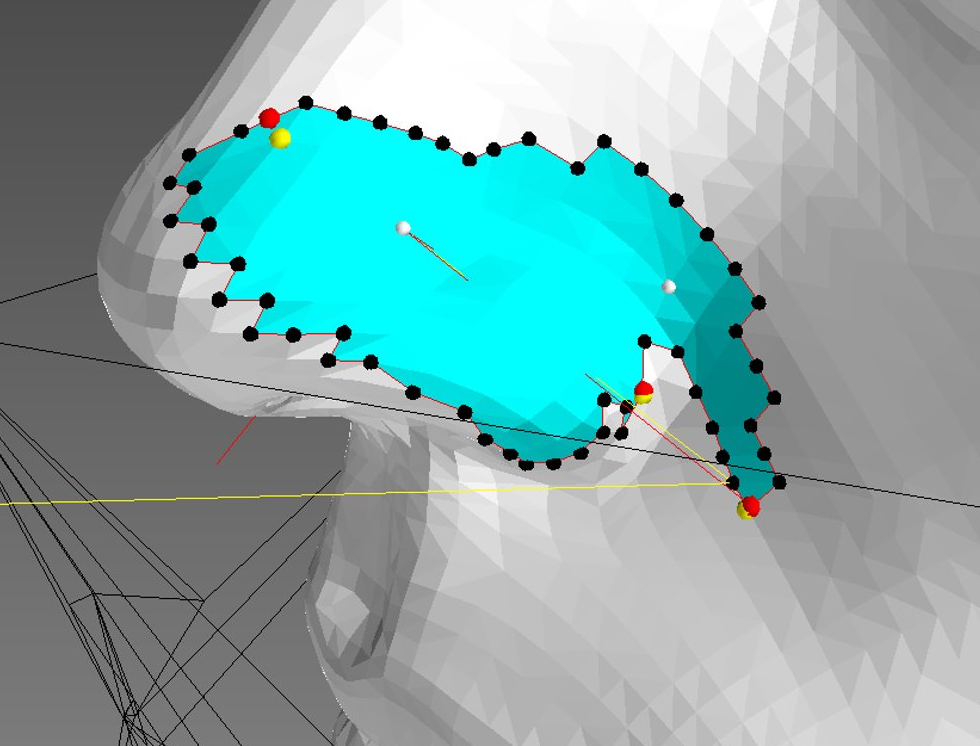}
\caption{Left: Fat pad movable border and handle. Middle: given a handle and a vertex, plane $P$ intersects the border. Right: nostril fat pad has a concave shape.}
\label{fig:intersection}
\end{figure}

\textbf{Attenuation Matrix.} A weight $w_{v,h}$ is computed for each pad's vertex $v$ and handle $h$, the closer the vertex to the border, the higher the attenuation. To avoid any border aberration, the weight must decrease slowly to zero at the border. We propose to use the following function that decreases according to the geodesic distance between the vertex position and the border,
\begin{equation}\label{function}
  w_{v,h}=\frac{(d(v,h)-d(i,h))^2}{d(i,h)^2},
\end{equation}
with $d$ the geodesic distance between two vertices, and $i$ the vertex located at the intersection of the border of the pad and the line $\vec{vh}$ (see Figure~\ref{fig:scheme} and \ref{fig:intersection}). The use of a quadratic radial basis function kernel appears appropriate as it smoothly goes from 1 (at the handle) to 0 (at the borders), thus restricting the influence on the fat pads. The function is used to compute the weight matrix $W_h$ for each vertex of each fat pad. This kernel has a zero derivative at the boundaries but it is not an issue in our case as we set the boundaries' vertices fixed and unmovable (weight is 0). 

\textbf{Specific Borders' Vertices.} Lips are a good example where fat pads must be processed differently.
Indeed, it has to follow the deformation of the fat pads to allow folding up, opening, and closing. Such borders can be specified in our model as an exception not to set the attenuation at the border to zero as displayed in Figure~\ref{fig:intersection}. It is the left upper lip fat pad with its two handles (blue) and its movable border (red). This is applied to the nostrils and the eyelids as well. In our posing process, we set the weight for this specific borders' vertices to 1, thus it moves freely according to the handle motion as if it was not linked to another adjacent pad.

The final positions $V'$ of vertices after moving a handle $h$ is:
\begin{equation}
  V' = V + (V_{gc} - V)W_h   
\end{equation}
with $V$ the current vertices' positions, $V_{gc}$ the GC positions and $W_h$ the weight matrix associated to the handle $h$.

\textbf{Pad Shape Processing.} As fat pads may have different shapes, the weight matrix $W_h$ must be computed for any new mesh. $d(i,h)$, distance between the handle and the border passing through vertex $v$, is a key value to compute $W_h$. It is different for each vertex and each handle. To compute $d(i,h)$, we set plane $P$, defined by line $L$ (between handle and vertex) and handle's normal $N$ (see Figure~\ref{fig:intersection}). 
$P$ may intersect the border at many points if the fat pad is non-convex (see Figure~\ref{fig:intersection}). Figure~\ref{fig:scheme} shows cases A, B and C on a non-convex fat pad with vertices $v_A$, $v_B$ and $v_C$, and handle $h$. Plane $P$ remains the same, and therefore the list of intersection points $l = [i_{A}, i_{B}, i_{C}, i_{F}]$. A way to remove intersection candidates is to filter according to direction: $(\widehat{\overrightarrow{hi}, \overrightarrow{hv}}) = 0$ (A) and then according to distance. A vertex $v_n$ must be between $h$ and $i_n$: $d(h,v)<d(h, i)$ (B and C). The min distance is kept $i_{select}|d(v,i_{select}) = min_{i\in l}(d(v, i))$. The weight matrix is computed offline using Surazhsky's algorithm for the geodesic distances~\cite{Surazhsky2005}.

\textbf{Cage Construction.} We select cages for two reasons. First, the shape preservation property of the cage combined with GC reduces undesired artifacts while preserving face volume. Second, the intuitiveness, simplicity of usage and direct manipulation given by cages suit our needs for non-experts. The cage is built using the position of the pad's handles. To get the cage topology, we compute a Delaunay triangulation on the handles, generating a tetrahedral mesh from which we extract the convex hull. This ensures a homogeneous connection with almost all the nearest neighbors. As we want to model facial poses, we need the face to be dynamic and the rest of the head to remain static. Thus, we do not want the cage to entirely encapsulate the head mesh but only the face. To build local cages several conditions are required~\cite{Lipman2008}: (1) the borders' vertices need to be fixed, (2) the cage must be scaled to prevent edges to intersect the model, and (3) the cage needs to be closed. To respect (1), cage borders' vertices are duplicated, scaled and then fixed. As cage vertices are based on pads' handles, fixing the initial borders' vertices do not make sense as it strongly limits the interaction with the pads. The duplication prevents this issue. To respect (2), cage's vertices are moved away from the mesh along the normal to ensure no intersection between the mesh and the cage. First, we apply a uniform scaling to the upper part of the face and a double uniform one to the lower part due to smaller and tightener number of handles on the jaws. Then, based on their position on the face, a specific non-uniform adjustment scaling is applied to some vertices. For example, nose handles are only scaled according to z-axis (towards the image plane) to keep them in the middle of the face. To respect (3), two new vertices are positioned at the back of the head of the model. GC are meant to affect all the vertices within the cage. For instance if the upper lip handle is moved up to open the mouth, and the lower lip handle is moved down, vertices of the lower lip are moved relatively to the new position of those of the upper lip. The expected behavior is that they should be independent. To tackle this, two cages are built, one for the upper part of the head and one for the lower one.

\section{Results}
\begin{figure}[h]
\centering
\includegraphics[width=\linewidth]{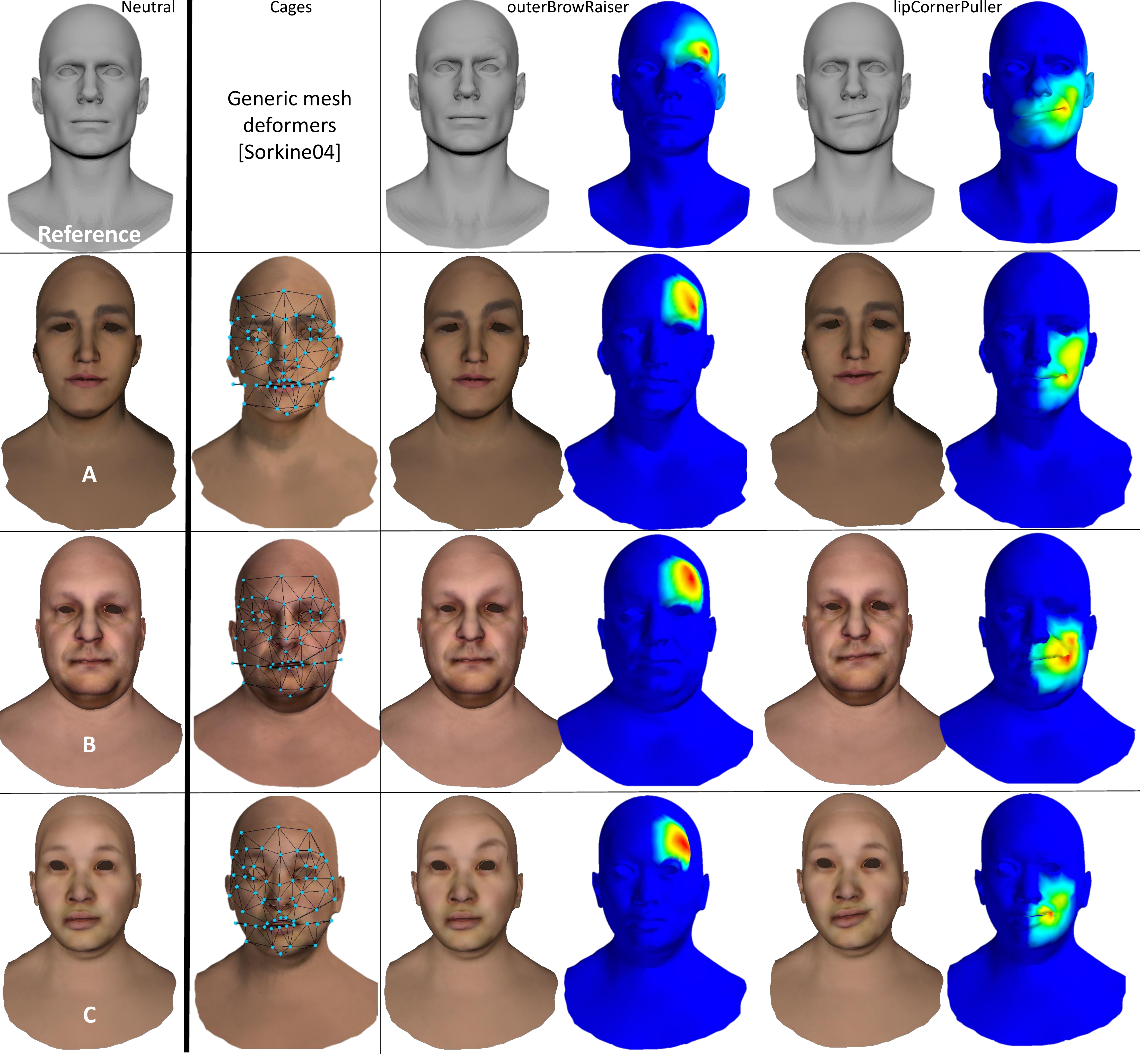}
\caption{Results of modeling two primary facial action units. References are made by an artist.}
\label{fig:blendShapes_cages_3people}
\end{figure}

Figure~\ref{fig:blendShapes_cages_3people} presents posing results, obtained by a non-expert on three different facial meshes, compared to a professional artist (top row). Only parts with movable handles (not the ones added to close the cage) are displayed. When a handle is grabbed, the associated fat pad is colored in light blue and the deformation center with a red sphere (actual handle's position on the pad). The artist used generic mesh deformers~\cite{Sorkine2004} whereas the non-expert has used our Fat Pad cage. The three meshes were obtained through a digital double creation pipeline~\cite{Danieau2019}. All meshes have the same mesh topology allowing the transfer of the template fat pads map. We selected two fundamental action units: AU-2 (Outer Brow Raiser) and AU-12 (Lip Corner Puller). They are involved in many facial human expressions (e.g. Happiness, Sadness, Fear etc.). Heat maps show differences between the initial mesh and the final modeling. Dark blue is static. We notice how close to the ground-truth the deformations are. It also emphasizes that our new Fat Pad cage prevents undesired mesh deformation due to global deformation (recomputation of the whole mesh when a cage vertex is moved) and ensures smooth deformation at the borders of the pads. With our cage, most of the mesh remains static, deformations only occur in the specific locations the user wanted to model.
\begin{figure}[h]
\centering
\includegraphics[width=\linewidth]{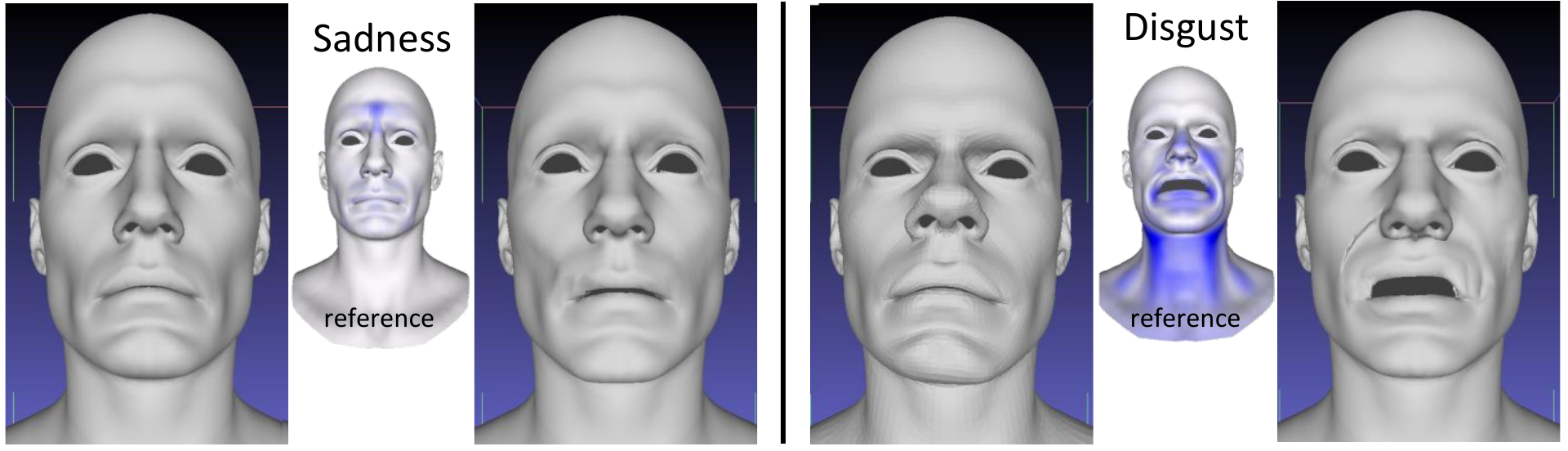}
\caption{Example of user results. From left to right: sadness (GC | ref | Fat Pad) and disgust (GC | ref | Fat Pad).}
\label{fig:resSad}
\end{figure}

\vspace{-0.5mm}
\section{Evaluation}
\begin{figure}[h]
\centering
\includegraphics[width=\linewidth]{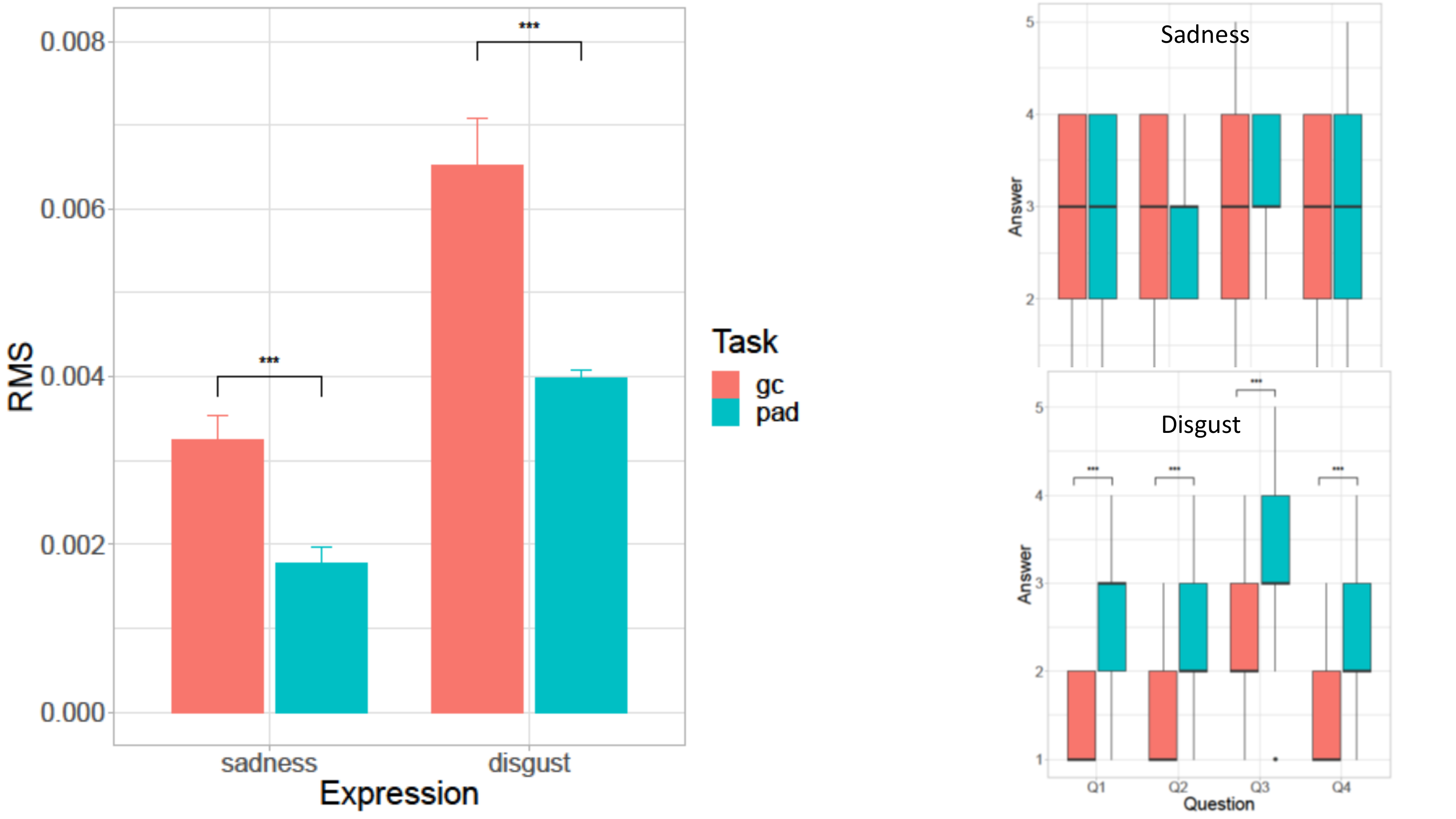}
\caption{Left: mean of the RMS of the Hausdorff distance between meshes. The lower the better. Right: answers for sadness (up) and disgust (bottom). The higher the better.}
\label{fig:rms}
\end{figure}

A user study was conducted with two cage systems: GC and Fat Pad. Two expressions were used as reference to reproduce: "sadness" and "disgust". With each cage participants had to realize the two expressions, leading to four tasks. We have chosen to compare our method to a single GC cage as used in the literature, and because a two-parts GC cage leads to strong artifacts around the mouth. By design, a local GC cage moves the whole mesh. Objective measures were time, number of handles' manipulation, number of undos, and the root mean square (RMS) of the Hausdorff distances between each participant's result and the reference model~\cite{hausdorff}. Subjective measures were the following questionnaire evaluated on a 5-point Likert scale, from 1 (totally disagree) to 5 (totally agree): [Q1] The result is similar to the model, [Q2] The task was easy, [Q3] The manipulation was easy, [Q4] I am happy with my result.  Participants had two screens, a front one displaying the interactive UI and a side one showing the reference meshes. It started with an exploration phase of 10 min to get used to the UI. The four tasks were then randomly presented. They stopped whenever they were satisfied with the result. Total duration was about 40 min. 17 participants took part in the study (age $\overline{x}= 33.75$ $\sigma = 12.920$). They were all used to 3D visualization and navigation ($\overline{x}= 3.65$ $\sigma = 1.069$ on a Likert scale 1 to 5) but no expertise in 3D modeling ($\overline{x}= 1.75$ $\sigma = 0.958$). A results sample can be seen on Figure~\ref{fig:resSad}. For the sadness expression, results produced with the Fat Pad cage are visually similar to the GC cage, both close to the reference. However, for the disgust expression, the Fat Pad cage is closer to the reference because of the possibility to open the mouth. Figure~\ref{fig:rms} shows the mean of the RMS of the Hausdorff distance between the participants' meshes and their reference. Meshes created with the Fat Pad are significantly closer to the reference than with the GC (Wilcoxon signed rank tests: $p=0.0021$ for sadness, and $p=3.05e^{-05}$ for disgust). However, no significant differences were found for other objective metrics.
Figure~\ref{fig:rms} shows answers to questionnaire. In the case of the sadness task, we did not observe statistical differences between the two cages ($p>0.05$). For a simple task, the performance with the Fat Pad is not perceived differently from with the GC. This is in line with the visual analysis (Figure~\ref{fig:resSad}). However, with the disgust task, all assertions were better rated for the Fat Pad (Q1: $p=0.001$, Q2: $p=0.001$, Q3: $p=0.03$, and Q4: $p=0.002$). This preference is also visible from the produced meshes, and was confirmed by the post-test interviews.

\section{Conclusions and Perspectives}
We presented Fat Pad cages, a first combination between cage-based deformation and facial anatomical model. The paper has described three main contributions: the Fat Pads concept enabling an interactive mesh deformation that respects facial anatomical constraints, a new automatic way of creating personalized cages for any facial mesh, and a user study validating the high interest and preference of Fat Pad cages. The new filter function appears to be more suitable to pose faces than Green Coordinates. It prevents global deformation and ensures smooth deformation at the borders of the pads. The generated cages closely fit the shape of the mesh and can be considered as an adaptive extension of the head. The user study confirmed the interest of our approach and provided valuable insight to improve our system. For instance, allowing the symmetrical control of the handles would be necessary to speed up the design of expressions. In a longer perspective, we want to focus our work to include machine learning to improve facial posing. Acquiring and analyzing facial fat motion would allow to better define pad's shape and regions of impact. We also want to extend the concept of multiple cages to other facial parts while keeping making facial posing easier and more accessible to non experts.



\bibliographystyle{ACM-Reference-Format}
\balance
\bibliography{biblio}

\end{document}